\documentclass[final,5p,times,twocolumn]{elsarticle}

\usepackage{amssymb,amsmath}

\usepackage{graphicx}
\usepackage{caption}
\usepackage{subcaption}
\usepackage{float}
\usepackage{bm}
\usepackage{color}
\usepackage[normalem]{ulem}
\usepackage{soul}
\journal{J. Non Newtonian Fluid Mech.}
\usepackage{algorithm} 
\usepackage{algpseudocode} 

\newcommand{\review}[1]{\textcolor{black}{#1}}

\newcommand{\Wi}{\text{Wi}}

\begin{document}

\begin{frontmatter}



\title{Reversal of particle Migration for viscoelastic solution at high solvent viscosity. }


\author[inst1,inst3]{Xavier Salas-Barzola}
\author[inst1]{Guillaume Maîtrejean}
\author[inst1]{Clément de Loubens}
\author[inst2]{Antoine Naillon}
\author[inst3,inst4]{Enric Santanach Carreras}
\author[inst1]{Hugues Bodiguel}

\affiliation[inst1]{organization={Univ. Grenoble Alpes, CNRS, Grenoble INP, LRP, 38000 Grenoble, France}}
\affiliation[inst3]{organization={TotalEnergies SE, Pole d'Etudes et Recherche de Lacq, BP 47-64170 Lacq, France}}
\affiliation[inst2]{organization={Univ. Grenoble Alpes, CNRS, Grenoble INP, 3SR, 38000 Grenoble, France}}
\affiliation[inst4]{organization={Laboratoire Physico-Chimie des Interfaces Complexes, Batiment CHEMSTARTUP, 64170 Lacq, France}}

\begin{abstract}
The imbalance of normal stress around a particle induces its transverse migration in pressure-driven viscoelastic flow, offering possibilities for particle manipulation in microfluidic devices. Theoretical predictions align with experimental evidence of particles migrating towards the center-line of the flow. However, these arguments have been challenged by both experimental and numerical investigations, revealing the potential for a reversal in the direction of migration for viscoelastic shear-thinning fluids. Yet, a significant property of viscoelastic liquids that remains largely unexplored is the ratio of solvent viscosity to the sum of solvent and polymer viscosities, denoted as $\beta$. We computed the lift coefficients of a freely flowing cylinder in a bi-dimensional Poiseuille flow with Oldroyd-B constitutive equations. 
A transition from a negative (center-line migration) to a positive (wall migration) lift coefficient was demonstrated with increasing $\beta$ values. Analogous to inertial lift, the changes in the sign of the lift coefficient were strongly correlated with abrupt (albeit small) variations in the rotation velocity of the particle. We established a scaling law for the lift coefficient that is proportional, as expected, to the Weissenberg number, but also to the difference in rotation velocity between the viscoelastic and Newtonian cases. If the particle rotates more rapidly than in the Newtonian case, it migrates towards the wall; conversely, if the particle rotates more slowly than in the Newtonian case, it migrates towards the center-line of the channel. Finally, experiments in microfluidic slits confirmed migration towards the wall for viscoelastic fluids with high viscosity ratio.
\end{abstract}




\end{frontmatter}


\section{Introduction}\label{secIntroduction}

Microfluidics has revolutionized the manipulation and characterization of particles at the microscale, enabling a wide range of applications in biology, chemistry, and materials science. Understanding the behavior of particles in microfluidic systems is crucial for optimizing device performance and achieving desired outcomes such as sorting and characterization of  particles. Following pioneer observations by Karnis and Mason \cite{karnis1966particle}, revisited in the context of the emergence of microfluidic device by Leshansky \textit{et al.}\cite{leshansky2007tunable}, the role of the viscoelasicity of the suspending fluid on particle dynamics has been investigated in details during the past decades (see for reviews references \cite{d2015particle, d2017particle, yuan2018recent, lu2017particle, zhou2020viscoelastic}). One of the most interesting features of confined viscoelastic flows is related to the existence of lift forces acting on solid particles, generally directed towards the center-line of the flow. These forces are very sensitive to the particle size, so that they could be used for particle sorting and separation applications. 

The viscoelastic lift force in a pressure driven flow originates from the normal stress \review{distribution} around the particle, which is unbalanced due to the fact that the shear rate is non uniform. This heuristic argument allows to write that the transverse elastic force $F_e$ scales as $F_e\sim r^3 \nabla N_1 $, where $r$ is the particle radius and \review{$\nabla N_1$ the gradient} of the first normal stress difference \cite{leshansky2007tunable,d2012single}. By balancing this force by the viscous drag, one expects that the migration velocity should be given by $V_M \sim \Wi z r^2/H^3$, where $\Wi$ is the Weissenberg number defined as the product of the viscoelastic relaxation  time by a mean shear rate, $H$ is the channel size (or tube diameter) and $z$ the distance to the mid-plane (or center-line). This scaling argument has been confirmed by analytical predictions \cite{ho1976migration,brunn1976slow}, numerical calculations \cite{d2012single} and experiments \cite{del2015effect, naillon2019dynamics}. 

Beyond this heuristic argument, a more nuanced understanding of the interplay between particles and viscoelastic shear-thinning fluids has emerged through extensive numerical simulations, shedding light on a complex picture, even in the absence of inertia.

Numerical investigations by Huang \textit{et al.} \cite{huang1997direct, huang2000effects} have explored the dynamics of particle migration in viscoelastic fluids exhibiting shear thinning. Contrary to the scaling argument presented above, particles in such fluids exhibit migration patterns that deviate from the traditional center-line orientation. Indeed, they can either migrate toward the center-line or the walls, resulting in the formation of an annular particle-free zone at intermediate radii. This result was obtained within the framework of the Oldroyd-B model coupled with the Carreau-Bird viscosity model in a two-dimensional pipe flow. Extending this exploration into three dimensions, D'Avino \textit{et al.} \cite{d2012single} computed the migration of particles in a pipe using the Giesekus model, wherein both the first and second normal stress differences ($N_1$ and $N_2$) were characterized by shear-thinning behavior. The results revealed that the direction of migration depends on both the initial position of the particles and the degree of shear-thinning. Experiments in square capillaries corroborated these results: particles migrated either towards the center-line or the corners of the capillary according to the degree of shear-thinning \cite{del2015effect}.

Constitutive models for viscoelastic fluids split the stress tensor into polymeric and solvent components, enabling the introduction of the viscosity ratio, denoted $\beta$, and defined as the ratio of the solvent viscosity to the sum of polymeric and solvent viscosity. Curiously, this parameter has been treated as a constant in numerical simulations \cite{huang1997direct, huang2000effects, d2012single, yu2019equilibrium}, despite its potential to vary by several orders of magnitude in experiments \cite{del2015effect} and microfluidic applications. This is probably due to the numerical computation challenges associated with the viscoelastic lift force remaining small at low Weissenberg numbers, requiring specific mesh refinements that are quite costly. Consequently, existing simulation outcomes are confined to a limited number of instances, and comprehensive parametric studies are currently unavailable.   

\review{In this paper, we demonstrate that the viscosity ratio of viscoelastic fluids plays a crucial role in determining the direction of migration of individual particles in Poiseuille flow. We employ the Olroyd-B constitutive model within a two-dimensional approximation of the geometry, allowing much faster calculations and a complete parametric study of the problem.  First, we outline the numerical methods used for computing the lift of a freely moving solid disk. 
Second, the impact of the Weissenberg number and viscosity ratio on the lift coefficient is shown. Finally, using experiments in a microfluidic slit, we confirm that increasing the viscosity ratio of a polymeric solution leads to a reversal in the direction of particle migration. }

It is important to note that, throughout this discussion, the term used to describe the object will be interchangeable, encompassing disk, cylinder, or particle. This interchangeability is contingent on the perspective, where geometrically it is viewed as a disk, numerically as a cylinder, and experimentally as a particle.

\section{Numerical model}

\subsection{Geometry and boundary conditions}
\begin{figure}[tb!]
  \centering
  \includegraphics[width=\linewidth]{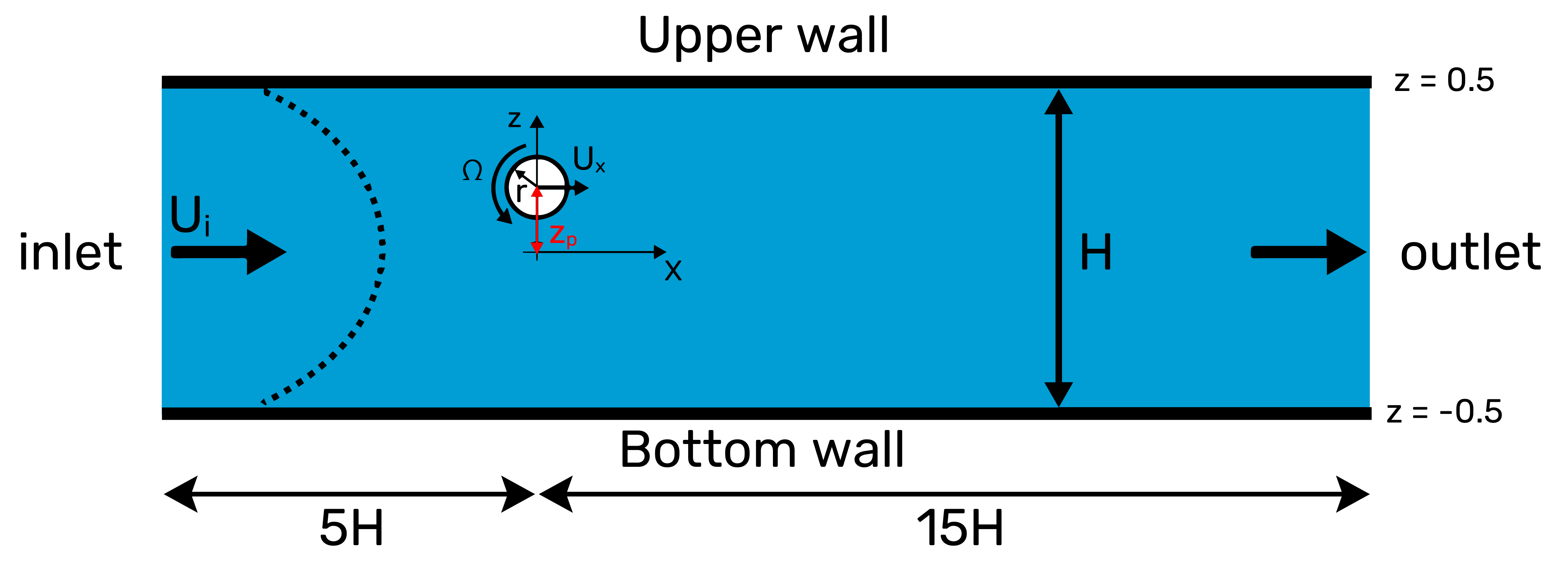}
    \caption{Problem definition of a rectangular domain of height H, length L=20H, with a free flowing disk, located at x=5H and z=$z_p$. $U_i$ is the inlet velocity of the plug flow, $\Omega$ and $U_x$ the rotational and translational velocity of the particle, respectively.}
    \label{figDomain}
\end{figure}

As depicted in Figure \ref{figDomain}, the geometry considered is a rectangular domain of height $H$ and length $L=20H$, in which a solid disk of diameter $d=2r=\alpha H$ was placed at a horizontal position of 5H from the inlet and a position $z_p H$ from the mid-plane. In this study, the confinement ratio $\alpha$ was set to 0.1 and was not varied. In the following, all the distances are made dimensionless by diving by the channel height $H$. The position $z_p$ was systematically varied between 0 and 0.35 as the problem is symmetric.

The flow is assumed steady and the Reynolds number very low, i.e. $Re \approx 0$. In that case, it is worth noting that the development length $\mathcal{L}$ of the flow to reach the parabolic velocity profile of the Hagen-Poiseuille flow is of the order of $H$, 
much smaller than the length 5H considered. The flow upstream the particle is thus fully developed.

The boundary conditions are depicted in Figure \ref{figDomain}. At the inlet, a plug flow velocity is applied, i.e. the velocity field writes $\bm{u} = \left( U_x, 0 \right)$, and an outflow condition is applied at the outlet. A no-slip condition is applied at both top and bottom walls.

\subsection{Governing equations}
\label{goveq}
The flow is simulated using OpenFOAM\textregistered, along with the RheoTool package \citep{rheoTool}, dedicated to solve non-Newtonian fluids, among others. \review{To enhance stability and convergence speed, the steady solution is achieved by obtaining a long-term solution to an unsteady problem, thereby retaining the time derivatives in the governing equations}. In the present case, the unsteady incompressible Stokes equations are solved for an Oldroyd-B fluid:



\begin{equation} \label{eqNS2}
  \partial_t\mathbf{u} = \frac{1}{\rho}\left(-\nabla p + \nabla\cdot \bm{\sigma} \right)
\end{equation}
with $\mathbf{u}$ the velocity field, $\rho$ the fluid density, $p$ the pressure and \review{$\bm{\sigma}$ the stress tensor which} is split into the solvent and polymeric stress $\bm{\tau}$
\begin{equation}
  \bm{\sigma} = 2 \eta_s \bm {D } + \bm{\tau}
\end{equation}
with $\eta_s$ the solvent viscosity, and $\bm {D }=[\nabla\mathbf{u} + (\nabla\mathbf{u})^T]/2$ the deformation tensor.

Considering an Oldroyd-B fluid, the polymeric stress writes:
\begin{equation}
  {\displaystyle \bm {\tau } +\lambda{\stackrel {\nabla }{\bm {\tau } }}=2\eta _{p}\mathbf {D} }
\end{equation}
with $\lambda$ the relaxation time of the fluid, $\eta_p$ the polymeric viscosity, and ${\stackrel {\nabla }{\bm {\tau }}}$ the upper-convected derivative of the extra-stress tensor $\bm {\tau }$:
\begin{equation}
  {\displaystyle {\stackrel {\nabla }{\bm {\tau }}}={\frac {\partial \bm {\tau }}{\partial t}}+\mathbf {u} \cdot \nabla \bm {\tau }-\nabla \mathbf {u} ^{T}\cdot \bm {\tau }-\bm {\tau }\cdot \nabla \mathbf {u} }
\end{equation}

At steady state, and in the limit of small Reynolds numbers, the above set of equations only involve two parameters, the solvent viscosity ratio $\beta$ ($= 1$ for a Newtonian fluid),
\begin{equation}
  \beta = \frac{\eta_s}{\eta_p+\eta_s}
 \label{EqBeta}
\end{equation}

and the Weissenberg number $Wi$,
\begin{equation}
  \Wi = \frac{3\lambda U_x}{2H} = \frac{\lambda u_{max}}{H}
 \label{EqWi}
\end{equation}
where $u_{max}$ is the maximum velocity in the midplane of the unperturbed Poiseuille flow. 

In order to minimize the numerical instabilities that can be observed in viscoelastic fluids simulations, the above equations are solved using a log-conformation tensor \citep{fattal2004constitutive}, using a semi-coupled solver ($p-\bm{u}$ coupled and $\bm{\tau}$ segregated).

\subsection{Free flowing particle algorithm}
\label{secFreeFlowingAlgo}

A neutrally buoyant spherical particle, that is freely advected in a circular or plane Poiseuille flow, lags behind the fluid velocity at a scale proportional to the \review{square} of its diameter, i.e. $\alpha^2$, and exhibits a rotational velocity $\Omega = 4z_p u_{max}/H$  \citep{kim2013microhydrodynamics}. In assessing the lift force experienced by a free-flowing particle using a stationary approach, we chose a static reference frame fixed with respect to the particle. In this context, it is necessary to impose boundary conditions to recreate a free-flowing state for the particle. Essentially, ensuring free-flowing conditions involves cancelling drag, lift and torque forces acting upon the particle. Here, we chose to cancel only drag and torque forces, in order to retrieve the lift force, by adapting both the x-component of the particle velocity $\bm{u}_p \cdot \bm{e}_x$ and its rotational velocity $\Omega$.   

The appropriate no-slip boundary conditions around the particle are linked to $ \bm{u}_p$ and $\Omega$. Determining these is not straightforward and cannot be known beforehand. To overcome this challenge, we used an iterative algorithm (see Algorithm \ref{algo1}) informed by a multivariate optimization approach; the \textit{scipy.optimize.root} function from the SciPy toolbox \citep{2020SciPy-NMeth}.

This iterative method finely adjusts the particle's rotational and $x$-translational velocities until both torque and drag forces, respectively, are counterbalanced. On reaching this equilibrium, only lift force remains as the dominant force acting on the particle.

\begin{algorithm}
\caption{Free-flowing algorithm to retrieve the lift coefficient and rotation velocity. $\mathcal{T}$ and $F_D$ are the torque and the drag force, respectively. }
\begin{algorithmic}[1]
\State $\bm{u}_p = \Omega = 0$ .
\While{$\mathcal{T}$ and $F_D > 10^{-12}$}
    \State Run simulation with $\bm{u}_p \cdot \bm{e}_x \neq 0$ and $\Omega \neq 0$  
    \State Get new $\mathcal{T}$ and $F_D$
    \State  \review{Adjust $\bm{u}_p \cdot \bm{e}_x$ and $\Omega $ } 
\EndWhile
\State Retrieving $F_L$
\end{algorithmic}
\label{algo1}
\end{algorithm}


Typically the \textit{hybrd} routine, serving as the method within the optimization algorithm, takes approximately 6 to 13 iterations to converge to the roots, i.e. cancelling both torque and drag forces acting on the particle. 


\subsection{Validation of the model}
\label{valmodel}
\begin{figure}[tb!]
  \centering
  \includegraphics[width=\linewidth]{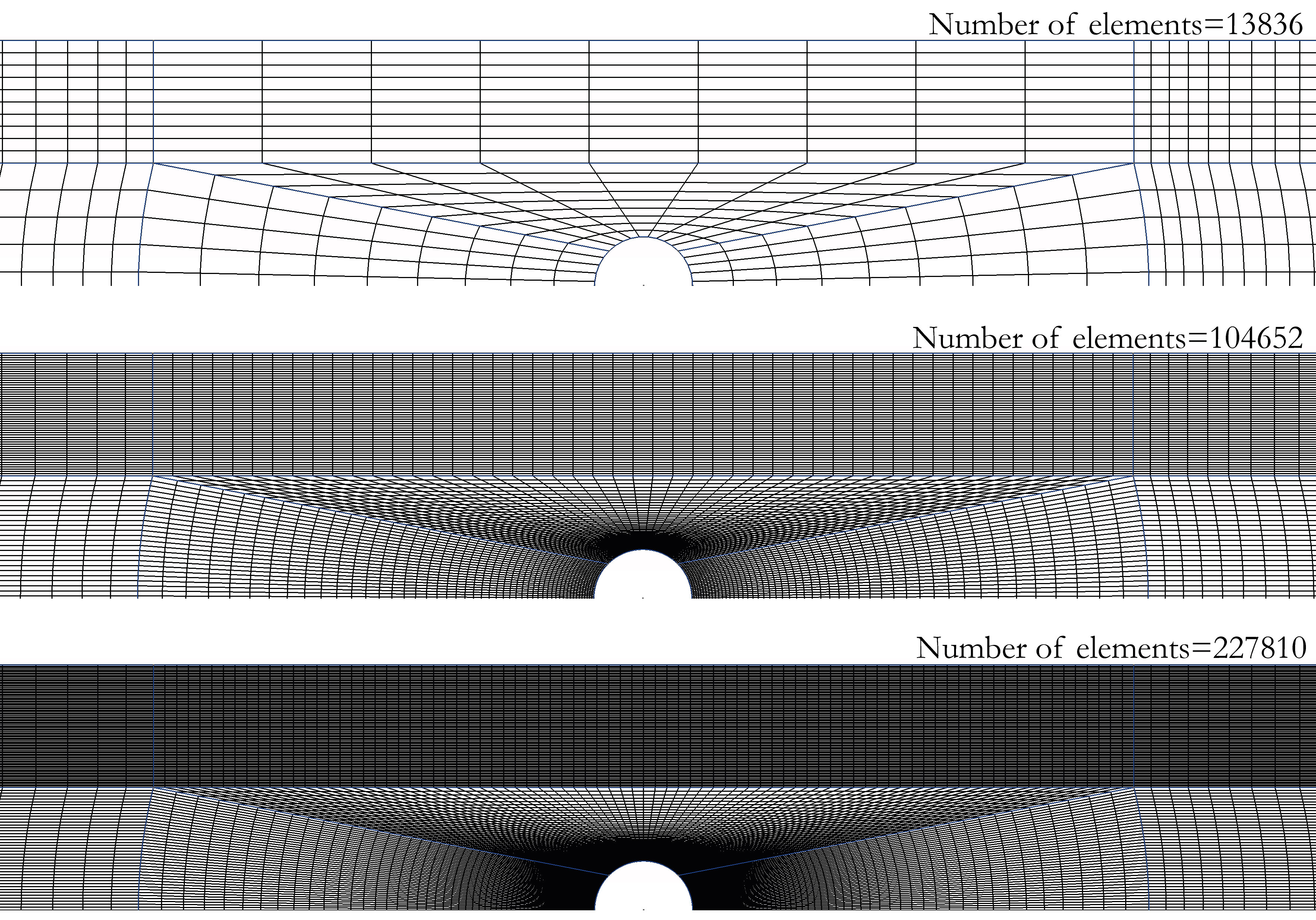}
  \caption{\review{Illustration of a subdomain centered on the disc, depicting a mesh generated using a block strategy with notable refinement around the particle boundary. The mesh for the entire domain, not depicted in this figure, comprises approximately 150k cells.}
  }
  \label{figMesh}
\end{figure}

Special attention was given to the meshing strategy of the domain, resulting in meshing a disk inside a rectangle using quadrangle cells. To do so, the domain was divided into boxes as pictured in Figure \ref{figMesh} with a significantly refined mesh generated in the so-called Region Of Interest (ROI), near the particle boundaries, using the mesh generator software GMSH \citep{geuzaine2009gmsh}. 

This meshing strategy has 2 major assets: the refinement is progressive and preserves the shape of the elements while the symmetry with respect to the x-axis is also preserved.

This last condition is very important when dealing with the across streamline migration, as the force $\bm{F}$ exerted on the particle:
\begin{equation}
  \bm{F} = \int_S{ \left( p\bm{I} + \bm{\sigma} \right)} \,dS
\end{equation}
is derived from the integral of the stress on the surface of the particle $S$ which can be many orders of magnitude lower than the drag force and thus, close to the numerical error caused by the mesh asymmetry.

It is also worth noting that the height of the ROI box is fixed and is arbitrarily set to two particle diameters and hence the maximum z-value of the particle center is $z_p=0.35$.

\begin{figure}[tb!]
  \centering
    \includegraphics[width=\linewidth]{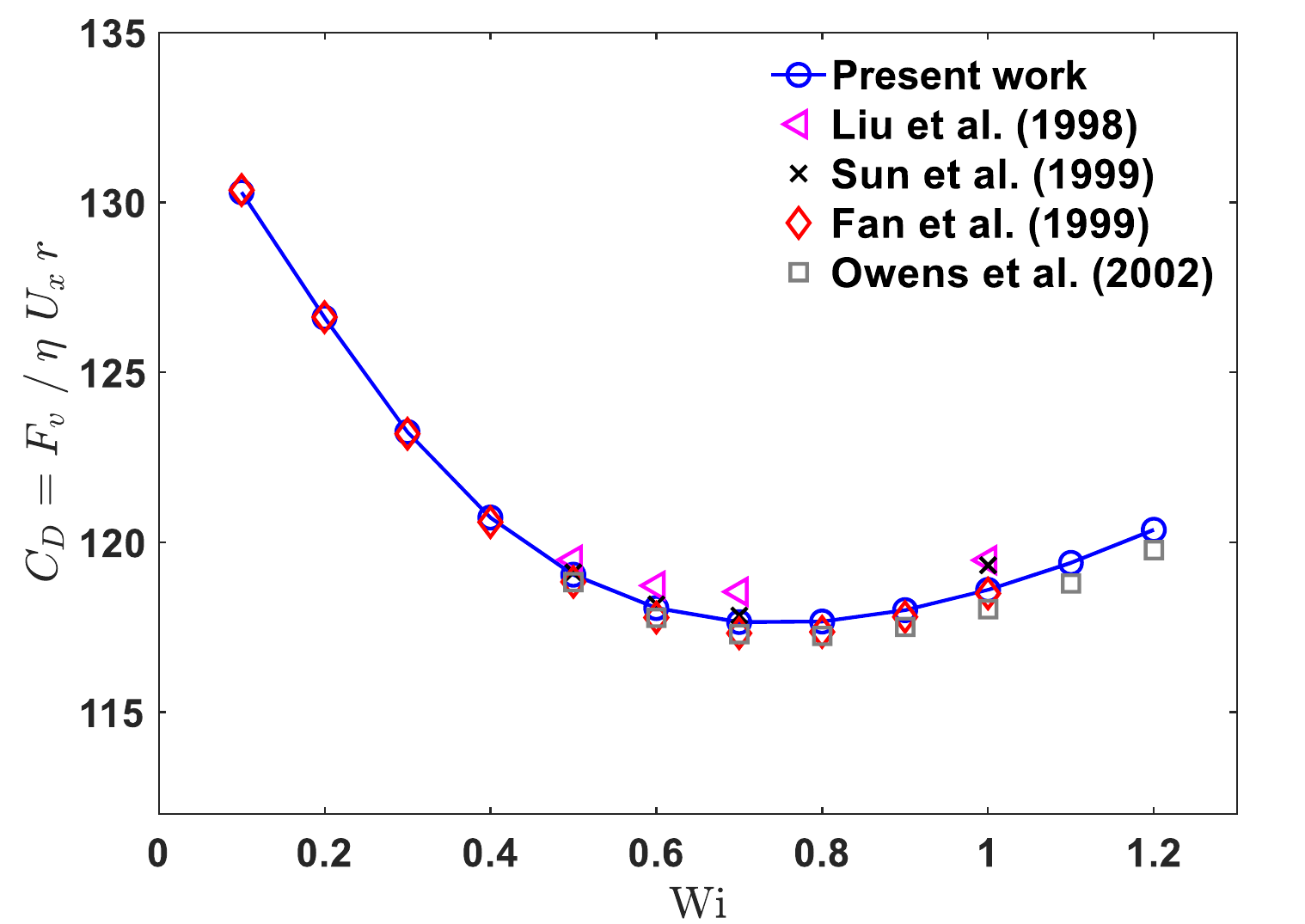}
  \caption{Drag coefficient experienced by a fixed cylinder in a planar flow of an Oldroyd-B fluid with $\alpha=0.5$, $\beta=0.59$, and $z_p=0$.}
  \label{figDragBench}
\end{figure}

Mesh convergence analyses were then performed and simulations involving the flow of a viscoelastic fluid around an fixed and confined cylinder were run to validate the numerical calculations. The literature abounds with reference results for Oldroyd-B fluid models in such geometry, facilitating a thorough comparison. Consistency with prior investigations was maintained by adopting a confinement ratio ($\alpha=0.5$), positioning the center of the disk at $z_p=0$, and specifying a solvent viscosity ratio of $\beta=0.59$, as specified in previous studies \cite{liu1998viscoelastic, sun1999finite, fan1999galerkin, owens2002locally}. The computed drag coefficient across Weissenberg numbers ranging from 0.1 to 1.2 was benchmarked against the established data and depicted in Figure \ref{figDragBench}, demonstrating excellent agreement.

\begin{figure}[tb!]
\centering
  \includegraphics[width=0.9\linewidth]{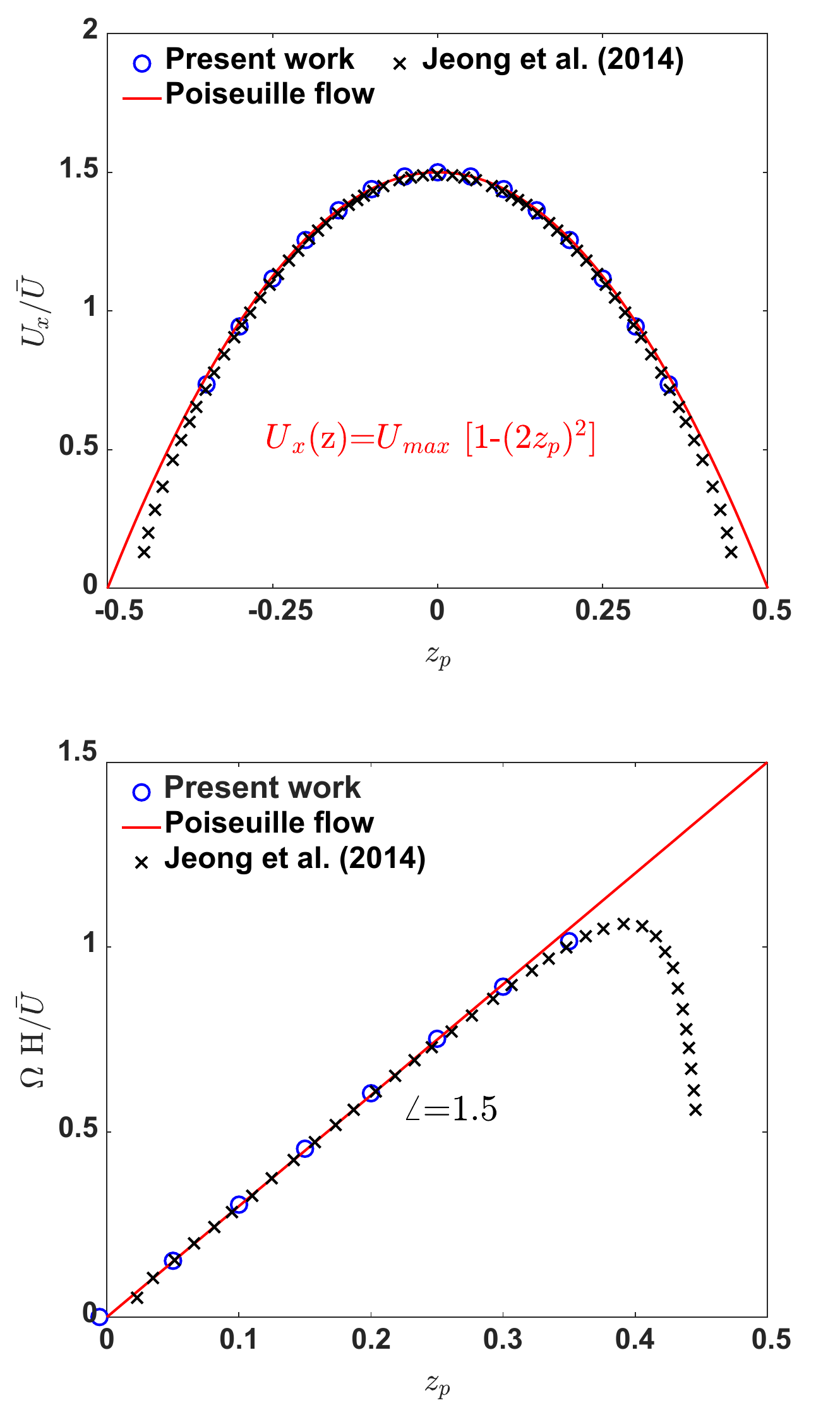}
  \caption{Longitudinal and rotational velocity of the particle in an Oldroyd-B fluid at low $\textrm{Wi}$ as a function of the $z$-position with $\alpha=0.1$,  $\beta=0.001$ and Wi=7.5 10$^{-4}$.}
  \label{figVelocitiesBench}
\end{figure}

Furthermore, Figure \ref{figVelocitiesBench} presents a comparison between the longitudinal and rotational velocities of the freely moving particle obtained in our study and the analytical solution proposed by \cite{jeong2014slow}. The analysis covers a z-position ranging from 0 to 0.35 with $\alpha=0.1$, $U_x=0.01$, $\beta=0.001$, and Wi = 7.5 10$^{-4}$. For a z-position close to the channel center, both longitudinal and rotational velocities are close to the ones observed in Newtonian Poiseuille flow, but deviated from them when moving away from the channel center. This deviation, also analytically observed by \cite{jeong2014slow}, was well caught by the present numerical model.

\section{Results and discussion}

\subsection{Lift coefficient at low $\beta$}

\begin{figure}[h!]
\centering
    \includegraphics[width=\linewidth]{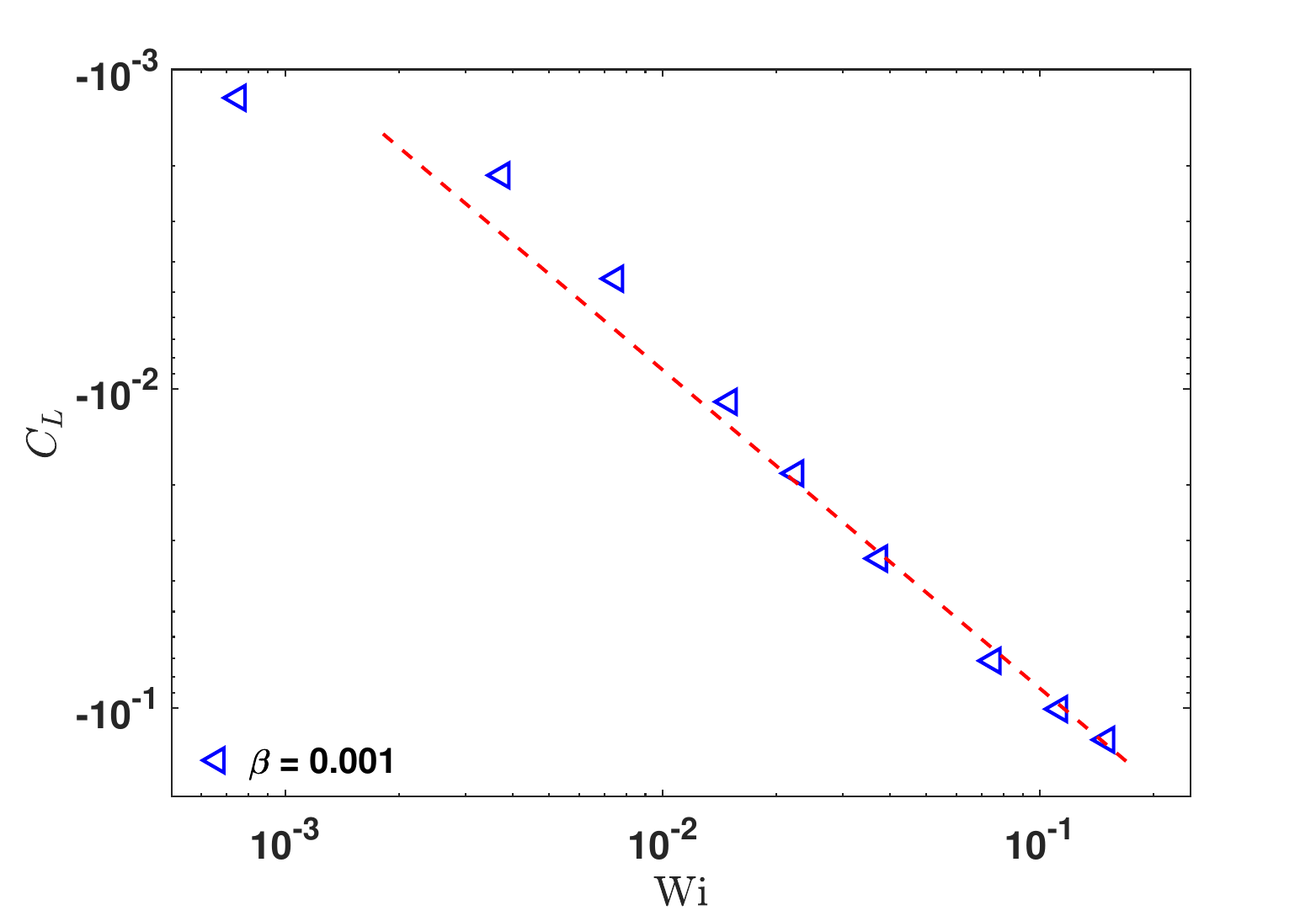}
    \includegraphics[width=\linewidth]{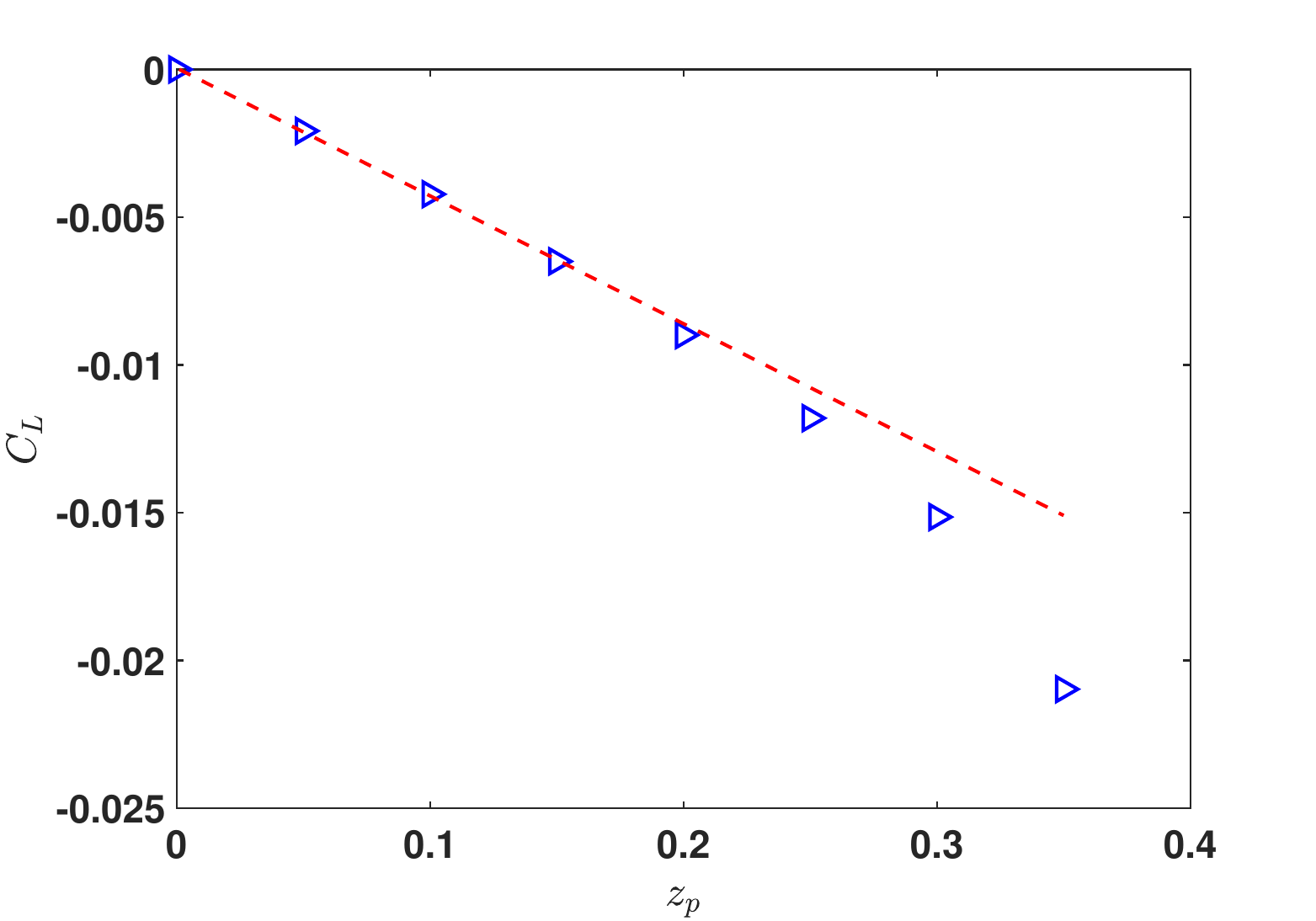}
  \caption{Lift coefficient for $\beta= 10^{-3}$. Top: as a function of $Wi$ at $z_{p}=0.15$. Bottom: as a function of $z_p$ for $Wi = 10^{-2}$. Dashed lines are the best linear fit to the data.} 
  \label{linearBlow}
\end{figure}

We first computed the lift coefficient $C_L$ in the low $\beta=\eta_s/(\eta_p+\eta_s)$ regime, i.e. for polymer solutions where the polymer viscosity is much greater than that of the solvent, and a fixed cylinder diameter ($\alpha = 0.1$). Figure \ref{linearBlow} shows $C_L$ as a function of the Weissenberg number $\text{Wi}$ and the position of the cylinder, $z_p$, for $\beta=10^{-3}$. 
In agreement with theoretical~\cite{ho1976migration}, numerical~\cite{d2017particle}, and experimental ~\cite{naillon2019dynamics} results obtained in the low $\beta$ limit, $C_L$ was negative (i.e., particles migrate towards the center) and proportional to $\text{Wi}$. It was also proportional to $z_p$ for $z_p<0.2$. Close to the wall, the variation of $C_L$ was more pronounced due to the proximity of the wall.

While it is not relevant to compare the order of magnitude of the lift forces between our 2D simulations and the previous 3D results, it is nevertheless worthwhile to verify that the lift velocities in both cases, which result from the transverse force, remain on the same order of magnitude in the low $\beta$ regime. In 3D, $C_L \simeq -15 \alpha ^2 \text{Wi} z$  \cite{naillon2019dynamics}, and the lift velocity  $U_z$ can be obtained by balancing the lift force with the drag force in the lateral direction, i.e. $U_z/U_x =  C_L/C_D \simeq 0.8 \text{Wi} \alpha^2 z$ (with $C_D = 6\pi$). In 2D, the drag coefficient is much higher, $C_D \simeq 130 $ for $\Wi = 0.1$, Figure \ref{figDragBench}. We found that the normalized lift velocity $U_z/U_x = C_L/C_D \simeq 3.3 \text{Wi} \alpha^2 z$. Therefore, the lift velocity of a cylinder (in 2D) is higher than that of a sphere in 3D, but remained of the same order of magnitude. In summary, at low $\beta$, we found a lift force in good agreement with previous works, proportional to $\text{Wi}$ and $z_p$, and directed towards the mid-plane of the channel.


\subsection{Reversal of direction of migration with $\beta$}

\begin{figure}[tb!]
\centering
    \includegraphics[width=\linewidth]{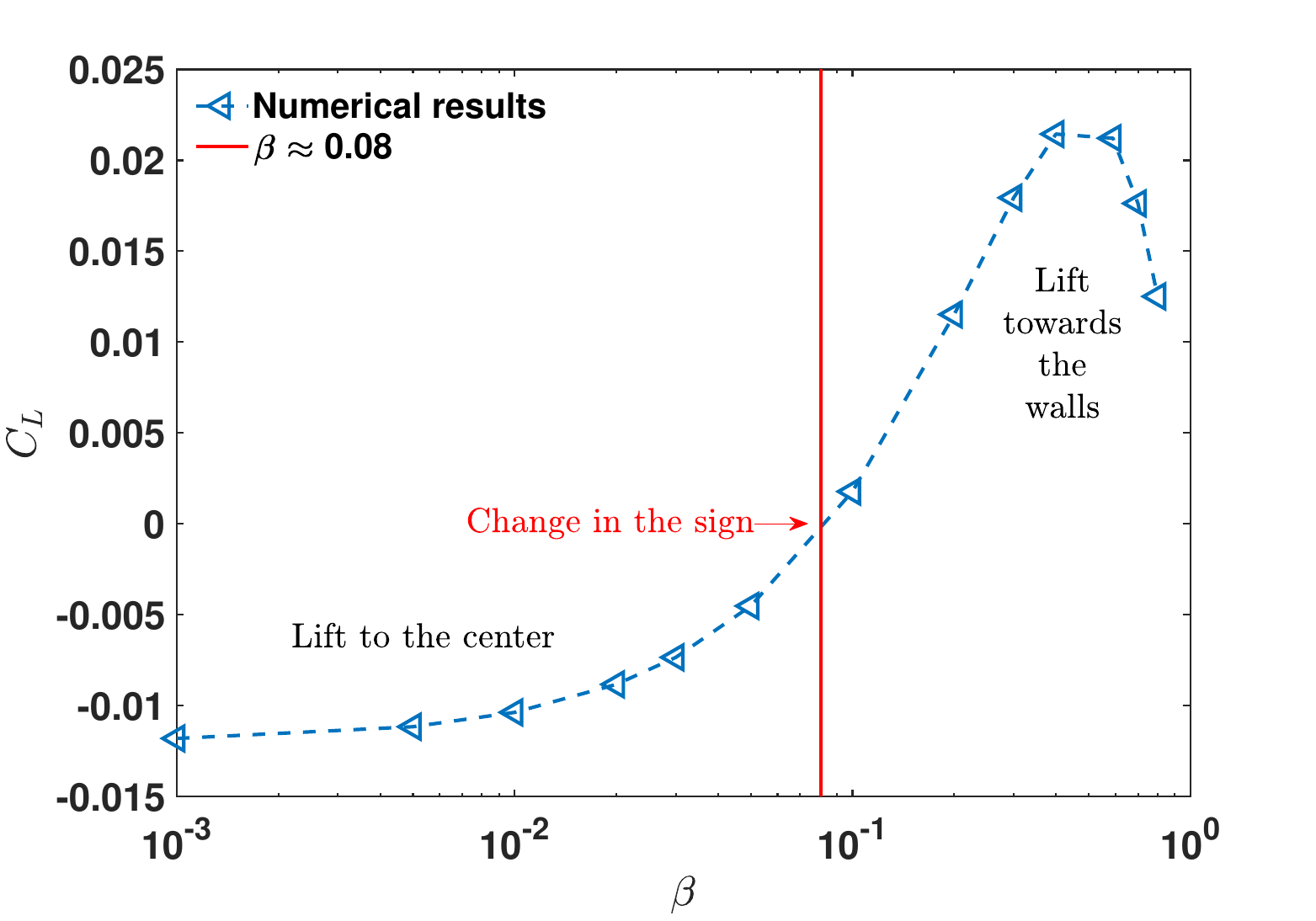}
    \caption{Lift coefficient as a function of $\beta$ for $Wi=0.01$, $z_p$=0.25.}
    \label{CLvsBeta}
\end{figure}

Surprisingly, we found that the direction of migration depends on the viscosity ratio $\beta$. Figure \ref{CLvsBeta} shows that $C_L$ undergoes a transition from negative to positive when $\beta$ exceeds 0.08, meaning that the disk migrates towards the wall instead of towards the mid-plane as was the case for lower values of $\beta$. For $\beta < 10^{-2}$, $C_L$ varies little with $\beta$. This asymptotic behavior derives directly from the Oldroyd-B model, since the contribution of the solvent to the stress is negligible as compared to the polymeric stress for low $\beta$. Around $\beta = 0.5$, the lift coefficient exhibited a maximum and then seemed to decrease towards zero when approaching unity. This limit is also expected when $\beta \rightarrow 1$, since the physical origin of the lift force lies in the polymeric stress, which becomes negligible in this limit.    

\begin{figure}[h!]
\centering
      \includegraphics[width=\linewidth]{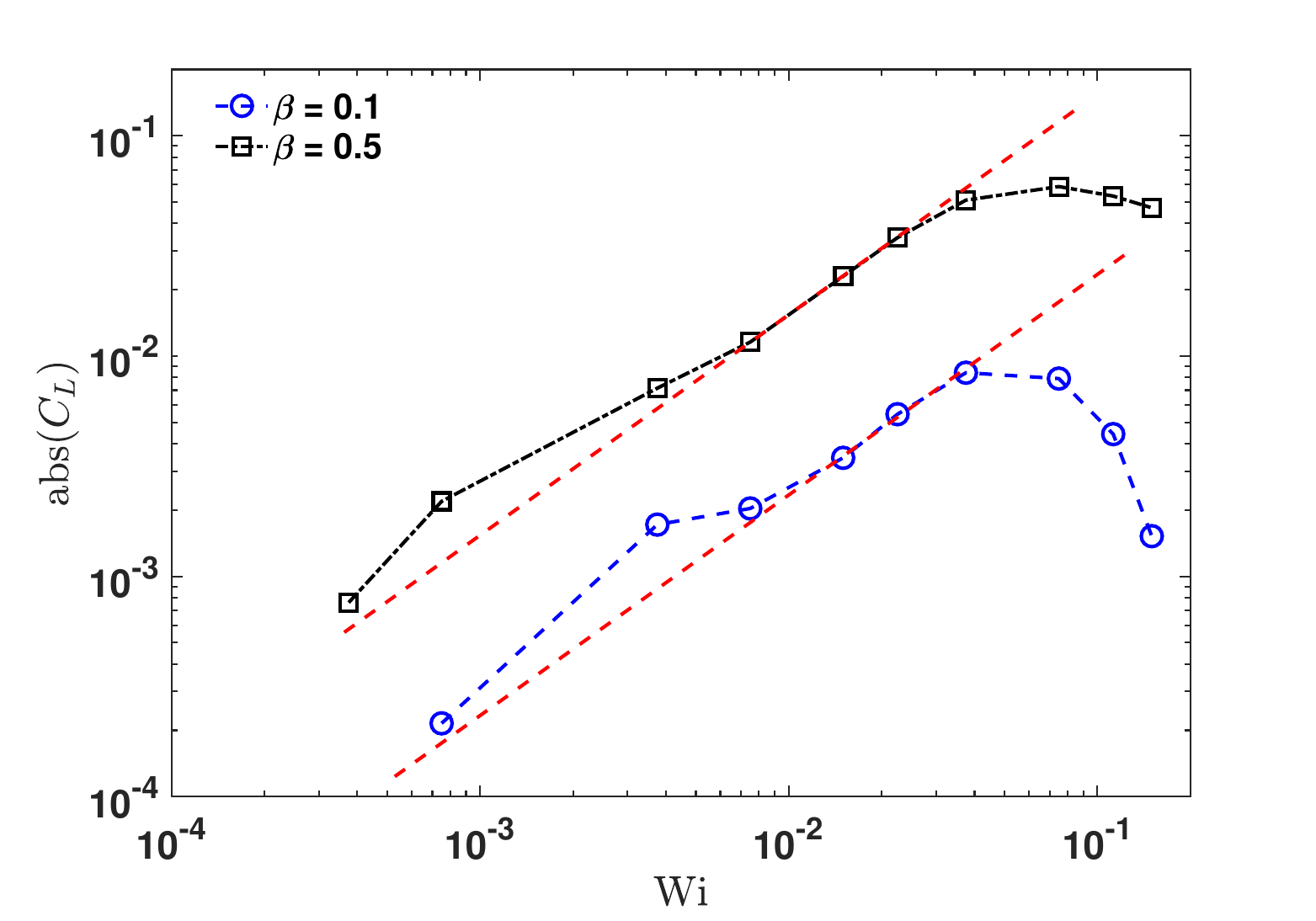}
      \includegraphics[width=\linewidth]{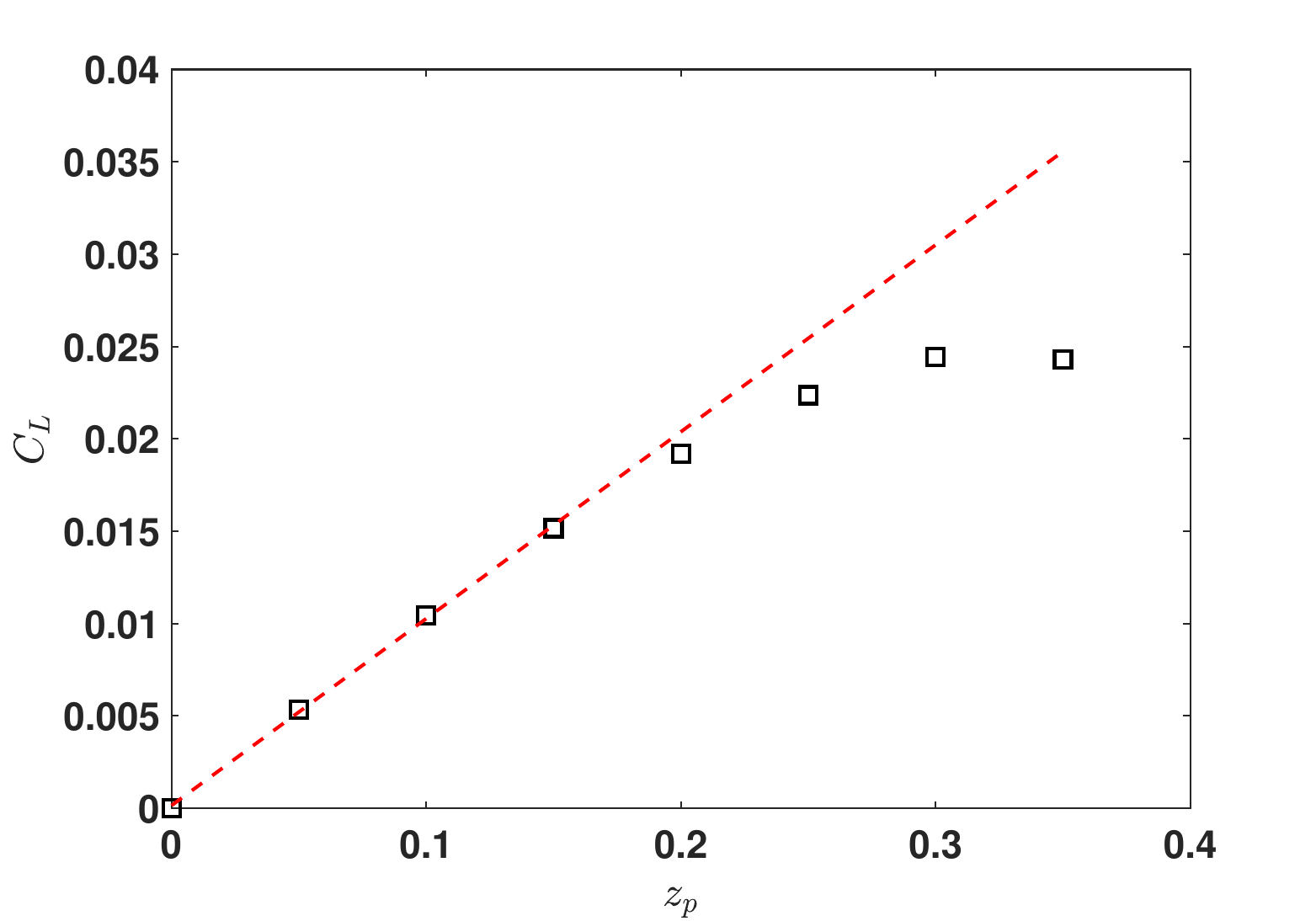}
    \caption{Top: lift coefficient versus $Wi$ for $\beta$=0.1 and 0.5 at $z_p=0.25$. Bottom: lift coefficient versus $z_p$ at $Wi=0.01$ and $\beta$=0.5. Dashed lines represent the best linear fits to the data. }
    \label{CLHighBeta}
\end{figure}

In the regime of positive lift (for $\beta>0.08$), the lift force, akin to the negative lift regime, exhibits a linear dependence on both $\Wi$ and $z_p$, especially for $\Wi<3\times 10^{-2}$ and for $z_p<0.2$, Figure \ref{CLHighBeta}. Consequently, the expression of $C_L$ as a function of $W_i$ and $z_p$ \citep{ho1976migration,d2017particle,naillon2019dynamics} can be generalized to low and high $\beta$ regime by considering that the prefactor $K$ depends on $\beta$, 
\begin{equation}
    C_L = K(\beta) \text{Wi} \alpha ^2 z_p
\end{equation}

The reversal of the migration direction with $\beta$ contrasts with phenomena arising from the shear-thinning properties of the viscoelastic liquid \citep{huang1997direct, huang2000effects, d2012single}. In this scenario, a critical position $z_p$ exists at which particles can migrate either towards the wall or towards the center-line of the flow.
In contrast, the direction of migration remains uniform regardless of $z_p$ variations when $\beta$ changes. Our results contrast the original analytical predictions for a second order fluid \cite{ho1976migration}, for which the non-Newtonian behaviour is limited to non-zero normal stress differences $N_1$ and $N_2$. \review{The direction of the lift could also be changed, as demonstrated in \cite{ho1976migration}, where it was found} that $ C_L \sim N_1-2N_2$. However, since in most cases $N_1 \gg N_2$ ($N_2=0$ for the Oldroyd-B model used here), the sign of $C_L$ remains negative. Both results therefore cannot be reconciled. 
 
This reversal of the direction of migration with $\beta$ is initially surprising, since the solvent viscosity ratio generally does not play a strong role in viscoelastic phenomena. However, let us mention the recent work of \cite{mokhtari_latche_quintard_davit_2022}, who emphasized the influence of the solvent viscosity ratio using the Oldroyd-B model, on the drag generated by a cylinder. At low $\beta$, stagnation zones were observed past the obstacles, due to the coupling between polymer conformation (viscoelastic stress) and the global stress field. These arrested zones were significantly smaller at high $\beta$. 

Overall, we could draw the conclusion that the value of the lift coefficient, and even its sign, strongly depends on the details of the rheological model used. 

\subsection{Reversal of direction of migration and particle rotational velocity}

\begin{figure}[tb!]
    \centering
     \includegraphics[width=0.95\linewidth]{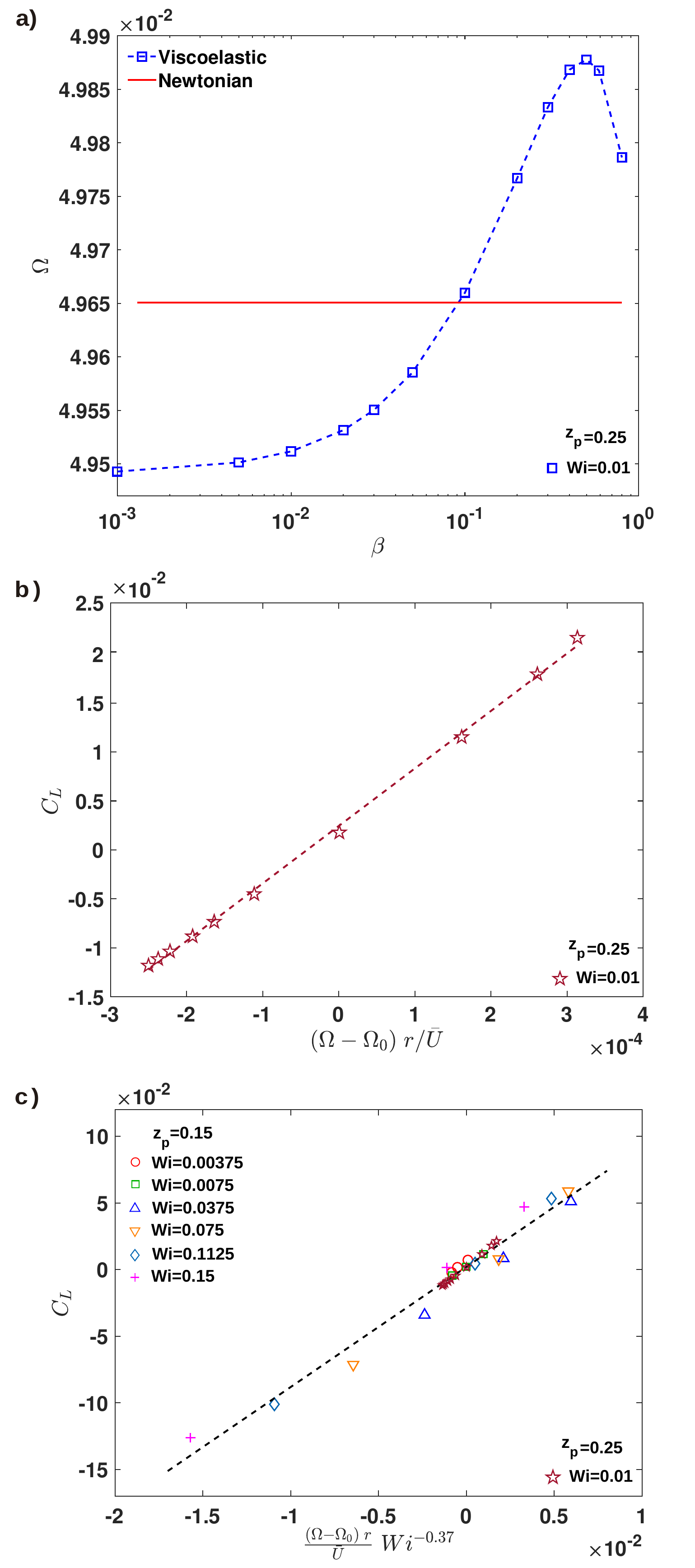}
    \caption{a) Particle rotational velocity $\Omega$ as a function of the viscosity ratio $\beta$, for $Wi=0.01$ and $z_p=0.25$. b) Lift coefficient $C_L$ as a function of the rotation velocity of the particle $\Omega$ for $\beta$ ranging from 0.001 to 0.8, $Wi = 0.01$ and $z_p=0.25$. c) Master-curve of $C_L$ as a function of $\left(\Omega - \Omega_0\right)\frac{r}{U} \text{Wi}^{-0.37}$ for different $Wi$ and $z_p$.}
    \label{fig:resumeContraintes2}
\end{figure}

To gain insights into the physical reasons for the change in sign of the lift coefficient with $\beta$, we conducted a detailed investigation of the perturbation of the velocity and stress fields induced by the rigid cylinder in various scenarios. Despite minor variations, conclusive findings could not be obtained. However, we found a strong correlation between the variations of $C_L$ and the angular velocity $\Omega$ with $\beta$. Figure \ref{fig:resumeContraintes2}-a shows a slight increase (less than 1\%) in $\Omega$ with $\beta$. Notably, for $\beta \simeq 0.1$, where the change of sign occurs, $\Omega$ is equal to that of the Newtonian case, resulting in no lift force. Higher angular velocities yield positive lift, while lower ones induce negative lift. 

This correlation between angular velocity and lift force is reminiscent of the inertial lift force \citep{joseph2002slip, yang2005migration}, for which it is well established that $C_L = f(\text{Re}) V_s \left(\Omega - \Omega_e\right)$ where $\Omega_e$ is the angular velocity of the particle (or cylinder) at the equilibrium transverse position, $V_s$ is the slip velocity, and $f(\text{Re})$ a pre-factor which depends on the Reynolds number. We adapted this concept to the viscoelastic lift by considering that the reference velocity of rotation $\Omega_e$ is given by the  Newtonian case. Figure \ref{fig:resumeContraintes2}-b shows that $C_L$ is proportional to $(\Omega - \Omega_0) \frac{r}{\bar{U}}$ for fixed $\text{Wi}$ and $z_p$. $C_L$ changes sign when $\Omega - \Omega_0$ also changes sign, which is when $\beta \simeq$ 0.08. We  found that the pre-factor is a power law of $\text{Wi}$, allowing us to write that

\begin{equation}
    C_L = 9 \left(\Omega - \Omega_0\right) \frac{r}{U} \text{Wi}^{-0.37}.
    \label{eqfinal}
\end{equation}

Figure \ref{fig:resumeContraintes2}-c shows the validity of this equation for two different positions of the particle ($z_p =$ 0.15 and 0.25) and $\text{Wi}$ ranging from $3.75 \times 10^{-3}$ to 0.15. The basis for this linear scaling of $C_L$ with $\Omega - \Omega_0$ can be justified by recalling that viscoelasticity breaks the symmetry of the Stokes equations. Consequently, changes in the rheology of the fluids modify the rotational velocity of the particle compared to the reversible Newtonian case. Depending on whether the particle rotates quickly or slowly, the migration direction will vary accordingly.

\subsection{Experimental evidence of migration towards the wall}

\begin{figure}[tb!]
    \includegraphics[width=0.90\linewidth]{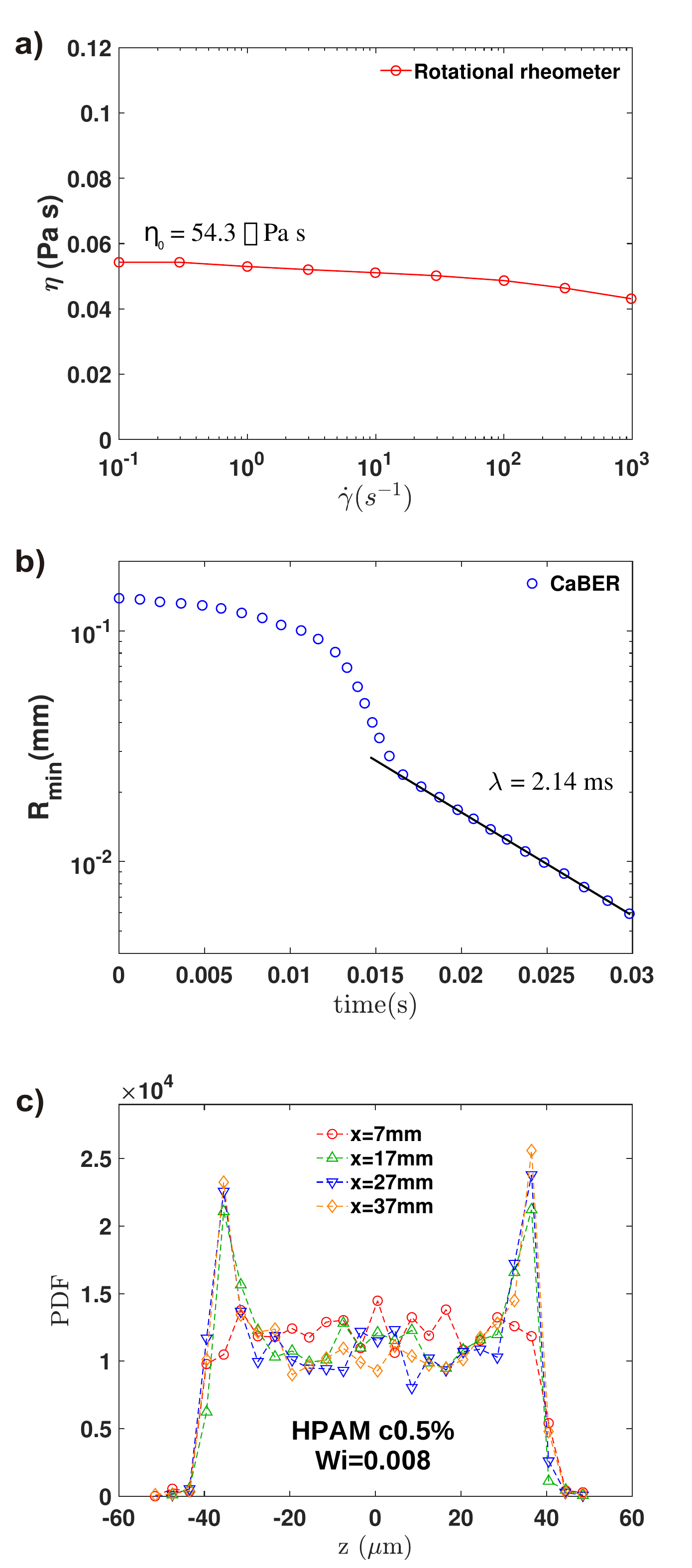}
    \caption{Experimental results obtained from HPAM 0.5\% w/w in a water-glycerol 30-70\% w/w solution, denoting a high solvent viscosity $\beta=0.4$. a) Flow curve obtained by rotational rheometry, cone-plate 1\textdegree: $\eta_0=54.3$ mPa.s. b) Relaxation time from capillary break-up experiment, 4 mm cylinder: $\lambda=2.14$ ms. c) Probability density functions (PDFs) of the $z$-position of 10 $\mu$m particles in a slit having a height of 100$\mu$m at Wi=0.008. $x$ is the distance from the inlet. \label{expval} }
\end{figure}

Finally, we attempted to observe experimentally the migration towards the walls in the high $\beta$ regime. We took advantage of the experimental set-up and methodologies presented in our previous publication \cite{naillon2019dynamics}, which were used to validate the scaling $C_L \simeq (2r/H)^2 \text{Wi} z $ in the low $\beta$ regime where the particles were focused in the median plane of the flow. Briefly, a diluted suspension of \review{neutrally buoyant} 10 $\mu$m diameter particles was injected under controlled pressure flow in a $100 \times 1000 $~$\mu$m slit of 5 cm length. The probability density function (PDF) of the $z$-position of solid particles at different travelling distances $x$ from the inlet was measured by fluorescent microscopy, see details in \cite{naillon2019dynamics}. Formulating the viscoelastic fluid for conducting experiments at high solvent viscosity posed significant challenges. We had to obtain a shear-rate independent shear viscosity, ensure that the fluid density matched that of the particles, and carefully select the appropriate quantity of solvent to maintain polymer solubility. Striking the right balance was crucial, as an excessively high solvent viscosity could result in extremely slow migration, requiring the use of excessively long channels and applied high pressure. Finally, we diluted a suspension of HPAM (CAS 9003-05-8) with an averaged molar mass of 150 kDa at 0.5\% in a 30/70 w/w water/glycerol mixture ($\rho=1190$ kg/m$^3$). The viscosity was determined by standard rheometry (cone-plate geometry on a TA DHR3 rheometer) and was found to be rate-independent, $\eta_0=54.3$ mPa s, $\beta =0.4$, Figure \ref{expval}-a. The relaxation time $\lambda$ was measured by capillary breakup experiment (CaBER) (see details in \citep{naillon2019dynamics}) and was $\lambda=2.14$ ms, Figure \ref{expval}-b.

The PDFs of the particle positions indicate an uniform distribution at the inlet ($x = 7$ mm), which slightly decreases in the center of the channel and doubles in regions close to the wall ($z = \pm 40$ $\mu$m) as we move further from the inlet, Figure \ref{expval}-c. This result clearly demonstrates that the particles migrated towards the wall in the presence of this viscoelastic fluid with a high solvent viscosity. The fact that the PDF remains relatively flat even far from the inlet is also consistent with the decrease in the lift force, approaching zero at the mid-plane of the channel. The absence of particles at the walls ($z=\pm 50$ $\mu$m) could be attributed to the significant interactions between the wall and the particles in the vicinity of the wall. However, this aspect was out of the scope of our numerical investigations.

\section{Conclusions}
We showed in the present study that the lift force, responsible for viscoelastic migration, had a complex dependency on rheological model parameters. In the limit of low solvent viscosity ratio - which represents the case of very long polymer chains - the results were in good agreement with the literature: the lift coefficient is proportional to the Weissenberg number and to the particle position. It is therefore proportional to the local shear rate of the unperturbed flow and leads to migration towards the center of the channel.

However, we have evidenced that the lift coefficient also depends significantly on the viscosity ratio $\beta$, decreasing when $\beta$ is increased and strikingly changing sign for $\beta>0.08$. For these high $\beta$ values, which correspond to very dilute polymer solutions in viscous solvents, the positive lift coefficient indicated a migration towards the wall. This dependency on $\beta$ clearly showed that the viscoelastic lift force was not simply given by the gradient of the first normal stress difference over the particle, but is sensitive to all the model parameters. Despite the physical reasons for this sign change of the lift coefficient remaining unclear, we noticed a robust linear correlation between the lift coefficient and the angular velocity difference with the Newtonian case, for a given Weissenberg number. The linear coefficient exhibited a decreasing power law variation as a function of Wi. This correlation resembles the one found for inertial migration in pressure-driven flow ($Wi=0$, $Re=O(1)$), where similarly, the lift coefficient is proportional to the angular velocity difference with that at equilibrium positions.

The change of sign of the lift force when $\beta$ is varied was qualitatively checked experimentally: a positive lift was found for a polymer solution having a high solvent viscosity ratio, leading to particle accumulation close to the walls of the microfluidic channel. This result indicates that although the numerical results were obtained in 2D, the above conclusions are valid in 3D, at least qualitatively. Their extension in 3D remains one of the most important perspectives of this work. Moreover, the apparent dependence of the lift on the details of the constitutive stress-strain relation of the system suggests a need for systematic studies involving more complex viscoelastic models.


\section*{Acknowledgments}
The Laboratoire Rh\'eologie et Proc\'ed\'es is part of the LabEx Tec 21 (Investissements d'Avenir - grant agreement n\textsuperscript{o}ANR-11-LABX-0030) and of the PolyNat Carnot Institut (Investissements d'Avenir - grant agreement n\textsuperscript{o}ANR-11-CARN-030-01).
\bibliographystyle{elsarticle-num} 
\bibliography{cas-refs}

\begin{thebibliography}{10}
\expandafter\ifx\csname url\endcsname\relax
  \def\url#1{\texttt{#1}}\fi
\expandafter\ifx\csname urlprefix\endcsname\relax\def\urlprefix{URL }\fi
\expandafter\ifx\csname href\endcsname\relax
  \def\href#1#2{#2} \def\path#1{#1}\fi

\bibitem{karnis1966particle}
A.~Karnis, S.~Mason, Particle motions in sheared suspensions. xix. viscoelastic media, Transactions of the Society of Rheology 10~(2) (1966) 571--592.

\bibitem{leshansky2007tunable}
A.~M. Leshansky, A.~Bransky, N.~Korin, U.~Dinnar, Tunable nonlinear viscoelastic “focusing” in a microfluidic device, Physical review letters 98~(23) (2007) 234501.

\bibitem{d2015particle}
G.~D’Avino, P.~L. Maffettone, Particle dynamics in viscoelastic liquids, Journal of Non-Newtonian Fluid Mechanics 215 (2015) 80--104.

\bibitem{d2017particle}
G.~D'Avino, F.~Greco, P.~L. Maffettone, Particle migration due to viscoelasticity of the suspending liquid and its relevance in microfluidic devices, Annual Review of Fluid Mechanics 49 (2017) 341--360.

\bibitem{yuan2018recent}
D.~Yuan, Q.~Zhao, S.~Yan, S.-Y. Tang, G.~Alici, J.~Zhang, W.~Li, Recent progress of particle migration in viscoelastic fluids, Lab on a Chip 18~(4) (2018) 551--567.

\bibitem{lu2017particle}
X.~Lu, C.~Liu, G.~Hu, X.~Xuan, Particle manipulations in non-newtonian microfluidics: A review, Journal of colloid and interface science 500 (2017) 182--201.

\bibitem{zhou2020viscoelastic}
J.~Zhou, I.~Papautsky, Viscoelastic microfluidics: Progress and challenges, Microsystems \& Nanoengineering 6~(1) (2020) 113.

\bibitem{d2012single}
G.~D'Avino, G.~Romeo, M.~M. Villone, F.~Greco, P.~A. Netti, P.~L. Maffettone, Single line particle focusing induced by viscoelasticity of the suspending liquid: theory, experiments and simulations to design a micropipe flow-focuser, Lab on a Chip 12~(9) (2012) 1638--1645.

\bibitem{ho1976migration}
B.~Ho, L.~Leal, Migration of rigid spheres in a two-dimensional unidirectional shear flow of a second-order fluid, Journal of Fluid Mechanics 76~(4) (1976) 783--799.

\bibitem{brunn1976slow}
P.~Brunn, The slow motion of a sphere in a second-order fluid, Rheologica Acta 15~(3-4) (1976) 163--171.

\bibitem{del2015effect}
F.~Del~Giudice, G.~D’Avino, F.~Greco, P.~A. Netti, P.~L. Maffettone, Effect of fluid rheology on particle migration in a square-shaped microchannel, Microfluidics and Nanofluidics 19 (2015) 95--104.

\bibitem{naillon2019dynamics}
A.~Naillon, C.~de~Loubens, W.~Ch{\`e}vremont, S.~Rouze, M.~Leonetti, H.~Bodiguel, Dynamics of particle migration in confined viscoelastic poiseuille flows, Physical Review Fluids 4~(5) (2019) 053301.

\bibitem{huang1997direct}
P.~Huang, J.~Feng, H.~H. Hu, D.~D. Joseph, Direct simulation of the motion of solid particles in couette and poiseuille flows of viscoelastic fluids, Journal of Fluid Mechanics 343 (1997) 73--94.

\bibitem{huang2000effects}
P.~Huang, D.~Joseph, Effects of shear thinning on migration of neutrally buoyant particles in pressure driven flow of newtonian and viscoelastic fluids, Journal of non-newtonian fluid mechanics 90~(2-3) (2000) 159--185.

\bibitem{yu2019equilibrium}
Z.~Yu, P.~Wang, J.~Lin, H.~H. Hu, Equilibrium positions of the elasto-inertial particle migration in rectangular channel flow of oldroyd-b viscoelastic fluids, Journal of Fluid Mechanics 868 (2019) 316--340.

\bibitem{rheoTool}
F.~Pimenta, M.~Alves, rheotool, \url{https://github.com/fppimenta/rheoTool} (2016).

\bibitem{fattal2004constitutive}
R.~Fattal, R.~Kupferman, Constitutive laws for the matrix-logarithm of the conformation tensor, Journal of Non-Newtonian Fluid Mechanics 123~(2-3) (2004) 281--285.

\bibitem{kim2013microhydrodynamics}
S.~Kim, S.~J. Karrila, Microhydrodynamics: principles and selected applications, Courier Corporation, 2013.

\bibitem{2020SciPy-NMeth}
P.~Virtanen, R.~Gommers, T.~E. Oliphant, M.~Haberland, T.~Reddy, D.~Cournapeau, E.~Burovski, P.~Peterson, W.~Weckesser, J.~Bright, S.~J. {van der Walt}, M.~Brett, J.~Wilson, K.~J. Millman, N.~Mayorov, A.~R.~J. Nelson, E.~Jones, R.~Kern, E.~Larson, C.~J. Carey, {\.I}.~Polat, Y.~Feng, E.~W. Moore, J.~{VanderPlas}, D.~Laxalde, J.~Perktold, R.~Cimrman, I.~Henriksen, E.~A. Quintero, C.~R. Harris, A.~M. Archibald, A.~H. Ribeiro, F.~Pedregosa, P.~{van Mulbregt}, {SciPy 1.0 Contributors}, {{SciPy} 1.0: Fundamental Algorithms for Scientific Computing in Python}, Nature Methods 17 (2020) 261--272.
\newblock \href {https://doi.org/10.1038/s41592-019-0686-2} {\path{doi:10.1038/s41592-019-0686-2}}.

\bibitem{geuzaine2009gmsh}
C.~Geuzaine, J.-F. Remacle, Gmsh: A 3-d finite element mesh generator with built-in pre-and post-processing facilities, International journal for numerical methods in engineering 79~(11) (2009) 1309--1331.

\bibitem{liu1998viscoelastic}
A.~W. Liu, D.~E. Bornside, R.~C. Armstrong, R.~A. Brown, Viscoelastic flow of polymer solutions around a periodic, linear array of cylinders: comparisons of predictions for microstructure and flow fields, Journal of Non-Newtonian Fluid Mechanics 77~(3) (1998) 153--190.

\bibitem{sun1999finite}
J.~Sun, M.~Smith, R.~Armstrong, R.~Brown, Finite element method for viscoelastic flows based on the discrete adaptive viscoelastic stress splitting and the discontinuous galerkin method: Davss-g/dg, Journal of Non-Newtonian Fluid Mechanics 86~(3) (1999) 281--307.

\bibitem{fan1999galerkin}
Y.~Fan, R.~I. Tanner, N.~Phan-Thien, Galerkin/least-square finite-element methods for steady viscoelastic flows, Journal of Non-Newtonian Fluid Mechanics 84~(2-3) (1999) 233--256.

\bibitem{owens2002locally}
R.~G. Owens, C.~Chauvi{\`e}re, T.~N. Philips, A locally-upwinded spectral technique (lust) for viscoelastic flows, Journal of non-newtonian fluid mechanics 108~(1-3) (2002) 49--71.

\bibitem{jeong2014slow}
J.-T. Jeong, C.-S. Jang, Slow motion of a circular cylinder in a plane poiseuille flow in a microchannel, Physics of Fluids 26~(12) (2014) 123104.

\bibitem{mokhtari_latche_quintard_davit_2022}
O.~Mokhtari, J.-C. Latché, M.~Quintard, Y.~Davit, Birefringent strands drive the flow of viscoelastic fluids past obstacles, Journal of Fluid Mechanics 948 (2022) A2.
\newblock \href {https://doi.org/10.1017/jfm.2022.565} {\path{doi:10.1017/jfm.2022.565}}.

\bibitem{joseph2002slip}
D.~Joseph, D.~Ocando, Slip velocity and lift, Journal of Fluid Mechanics 454 (2002) 263--286.

\bibitem{yang2005migration}
B.~H. Yang, J.~Wang, D.~D. Joseph, H.~H. Hu, T.-W. Pan, R.~Glowinski, Migration of a sphere in tube flow, Journal of Fluid mechanics 540 (2005) 109--131.

\end{thebibliography}

\end{document}